\documentclass[pra, showpacs, twocolumn, a4paper, nofootinbib]{revtex4}
\usepackage{graphicx}
\usepackage{amsmath, amsfonts, amssymb, bm}
\usepackage{amsmath}
\usepackage{verbatim}
\usepackage{bm}

\begin{document}

\newcommand{\ket}[1]{| #1 \rangle} \newcommand{\bra}[1]{\langle #1 |}

\title{Dark periods and revivals of entanglement in a two qubit
  system} \author{Z. \surname{Ficek}$^{a}$}
\author{R. \surname{Tana\'{s}}$^{b}$}
\affiliation{$^{a}$Department of Physics, School of Physical Sciences,
  The University of Queensland, Brisbane, Australia 4072\\ 
  $^{b}$Nonlinear Optics Division, Institute of Physics, Adam
  Mickiewicz University, Pozna\'n, Poland}

\date{\today}

\begin{abstract}
  In a recent paper Yu and Eberly [Phys. Rev. Lett. {\bf 93}, 140404
  (2004)] have shown that two initially entangled and afterwards not
  interacting qubits can become completely disentangled in a finite
  time. We study transient entanglement between two qubits coupled
  collectively to a multimode vacuum field and find an unusual feature
  that the irreversible spontaneous decay can lead to a revival of the
  entanglement that has already been destroyed. The results show that
  this feature is independent of the coherent dipole-dipole
  interaction between the atoms but it depends critically on whether
  or not the collective damping is present. We show that the ability
  of the system to revival entanglement via spontaneous emission
  relies on the presence of very different timescales for the
  evolution of the populations of the collective states and coherence
  between them.
\end{abstract}

\pacs{03.67.Mn, 42.50.Fx, 42.50.Lc}

\maketitle


The problem of controlling the evolution of entanglement between atoms
(or qubits) that interact with the environment has received a great
deal of attention in recent years~\cite{phbk,ft02,ms04,sm06}. The
environment may be treated as a reservoir and it is well
known that the interaction of an excited atom with the reservoir
leads to spontaneous emission that is one of the major sources of
decoherence. In light of the experimental investigations, the
spontaneous emission leads to irreversible loss of information encoded
in the internal states of the system and thus is regarded as the main
obstacle in practical implementations of entanglement.

This justifies the interest in finding systems where the spontaneous
emission is insignificant. However, in many treatments of the
entanglement creation and entanglement dynamics, the coupling of atoms
to the environment is simply ignored or limited to the interaction of
the atoms with a single mode cavity~\cite{zg00,os01}.

It is well known that under certain circumstances a group of atoms can
act collectively that the radiation field emitted by an atom of the
group may influence the dynamics of the other
atoms~\cite{dic,le70,ag74,fs05}. The resulting dynamics and the
spontaneous emission from the atoms may be considerably modified. It
was recently suggested that two suitably prepared atoms can be
entangled through the mutual coupling to the vacuum
field~\cite{phbk,ft02,bd00,ft03}.

It has also been predicted that two initially entangled and afterwards
not interacting atoms can become completely disentangled in a time
much shorter than the decoherence time of spontaneous emission. This
feature has been studied by Yu and Eberly~\cite{ye04} and
Jak\'{o}bczyk and Jamr\'{o}z~\cite{ja06}, who termed it as the "sudden
death" of entanglement, and have elucidated many new characteristics of
entanglement evolution in systems of two atoms. Their analysis,
however, have concentrated exclusively on systems of independent
atoms.

In this paper, we consider a situation where the atoms are coupled to
the multimode vacuum field and demonstrate the occurrence of multiple
dark periods and revivals of entanglement induced by the irreversible
spontaneous decay. We fully incorporate collective interaction between
the atoms and study in detail the dependence of the revival time on
the initial state of the system and on the separation between the
atoms. We emphasize that the revival of entanglement in a pure
spontaneous emission process contrasts the situation of the coherent
exchange of entanglement between atoms and a cavity
mode~\cite{zg00,os01}.

We consider two identical two-level atoms (qubits) having lower levels
$|g_{i}\rangle$ and upper levels $|e_{i}\rangle$ ($i=1,2$) separated
by energy $\hbar\omega_{0}$, where $\omega_{0}$ is the transition
frequency. The atoms are coupled to a multimode radiation field whose
the modes are initially in the vacuum state $\ket{\{0\}}$. The atoms
radiate spontaneously and their radiation field exerts a strong
dynamical influence on one another through the vacuum field modes.
The time evolution of the system is studied using the
Lehmberg--Agarwal~\cite{le70,ag74,fs05} master equation, which
reads~as
\begin{eqnarray}
  \frac{\partial \rho}{\partial t} &=& -i\omega_{0}\sum_{i=1}^{2} \left[S^{z}_{i},\rho\right]
  -i\sum_{i\neq j}^{2}\Omega_{ij}\left[S^{+}_{i}S^{-}_{j},\rho\right] \nonumber \\
  &-& \frac{1}{2}\sum_{i,j=1}^{2}\gamma _{ij}\left( \left[\rho S_{i}^{+},S_{j}^{-}\right]
    +\left[S_{i}^{+},S_{j}^{-}\rho\right]\right) ,\label{e1}
\end{eqnarray}
where $S_{i}^{+}\ (S_{i}^{-})$ are the dipole raising (lowering)
operators and $S^{z}$ is the energy operator of the $i$th atom,
$\gamma_{ii}\equiv \gamma$ are the spontaneous decay rates of the
atoms caused by their direct coupling to the vacuum field. The
parameters $\gamma_{ij}$ and $\Omega_{ij}\ (i\neq j)$ depend on the
distance between the atoms and describe the collective damping and the
dipole-dipole interaction defined, respectively, by
\begin{eqnarray}
  \gamma_{ij} = \frac{3}{2}\gamma\left( \frac{\sin  kr_{ij}}{kr_{ij}}
    +\frac{\cos  kr_{ij}}{\left( kr_{ij}\right) ^{2}}-\frac{\sin kr_{ij} }
    {\left( kr_{ij}\right) ^{3}}\right)  ,\label{e2}
\end{eqnarray}
and
\begin{eqnarray}
  \Omega_{ij} = \frac{3}{4}\gamma\left(-\frac{\cos  kr_{ij}}{kr_{ij}}
    +\frac{\sin kr_{ij} }{\left( kr_{ij}\right)^{2}}+\frac{\cos kr_{ij} }
    {\left( kr_{ij}\right) ^{3}}\right) ,\label{e3}
\end{eqnarray}
where $k =\omega_{0}/c$, and $r_{ij}
=\left|\vec{r}_{j}-\vec{r}_{i}\right|$ is the distance between the
atoms. Here, we assume, with no loss of generality, that the atomic
dipole moments are parallel to each other and are polarized in the
direction perpendicular to the interatomic axis. The effect of the
collective parameters on the time evolution of the entanglement in the
system is the main concern of this paper.

It will prove convenient to work in the basis of four collective
states, so-called Dicke states, defined as~\cite{dic}
\begin{eqnarray}
  |e\rangle &=& |e_{1}\rangle\otimes|e_{2}\rangle ,\nonumber \\
  |g\rangle &=& |g_{1}\rangle\otimes|g_{2}\rangle  ,\nonumber \\
  |s\rangle &=& \left(\ket{g_{1}}\otimes\ket{e_{2}}
    +\ket{e_{1}}\otimes\ket{g_{2}}\right)/\sqrt{2} ,\nonumber \\ 
  |a\rangle &=& \left(\ket{g_{1}}\otimes\ket{e_{2}}
    -\ket{e_{1}}\otimes\ket{g_{2}}\right)/\sqrt{2} . \label{e4}
\end{eqnarray}
In this basis, the two-atom system behaves as a single four-level
system with the ground state $\ket g$, two intermediate states $\ket
s$ and $\ket a$, and the upper state $\ket e$. As a result, the
problem of entanglement evolution in the two qubit system can be
determined in terms of populations and coherences between the
collective levels.

In order to determine the amount of entanglement between the atoms and
the entanglement dynamics, we use concurrence that is the widely
accepted measure of entanglement. The concurrence introduced by
Wootters~\cite{woo} is defined as
\begin{eqnarray}
  {\cal C}(t) =
  \max\left(0,\sqrt{\lambda_{1}}-\sqrt{\lambda_{2}}-\sqrt{\lambda_{3}}
    -\sqrt{\lambda_{4}}\,\right) ,\label{3.11} 
\end{eqnarray}
where $\{\lambda_{i}\}$ are the the eigenvalues of the matrix
\begin{equation}
  R=\rho\tilde{\rho} ,\quad {\rm with}\quad  \tilde{\rho} = 
  \sigma_{y}\otimes\sigma_{y}\,\rho^{*}\,\sigma_{y}\otimes\sigma_{y} ,\label{3.12}  
\end{equation}
and $\sigma_{y}$ is the Pauli matrix. The range of concurrence is from
0 to 1. For unentangled atoms ${\cal C}(t)=0$ whereas ${\cal C}(t)=1$ for
the maximally entangled atoms.

The density matrix, which is needed to calculate ${\cal C}(t)$ is
readily evaluated from the master equation~(\ref{e1}).  Following Yu
and Eberly, we choose the atoms to be at the initial time $(t=0)$
prepared in an entangled state of the form
\begin{eqnarray}
  \ket{\Psi_{0}} = \sqrt{p}\,\ket e + \sqrt{1-p}\,\ket g ,\label{ea}
\end{eqnarray}
where $0\leq p\leq 1$. The state $\ket{\Psi_{0}}$ is a linear
superposition of only those states of the system in which both or
neither of the atoms is excited. As discussed in
Refs.~\cite{ye04,ja06}, in the absence of the coupling between the
qubits, the initial entangled state of the form (\ref{ea})
disentangles in a finite time. They termed this feature as the sudden
death of entanglement.

In what follows, we examine the time evolution of the entanglement of
two atoms coupled to the multimode vacuum field.  If the atoms
are initially prepared in the state (\ref{ea}), it is not difficult to
verify that the initial one-photon coherences are zero, i.e.
$\rho_{es}(0)=\rho_{ea}(0)=\rho_{sg}(0)=\rho_{ag}(0)=\rho_{as}(0)=0$.
Moreover, the coherences remain zero for all time, that they cannot be
produced by spontaneous decay. This implies that for all times, the
density matrix of the system represented in the collective basis (\ref{e4}), 
is given in the block diagonal form
\begin{eqnarray}
  \rho(t) = \left(
    \begin{array}{cccc}
      \rho_{ee}(t) & \rho_{eg}(t) & 0 & 0 \\
      \rho_{eg}^{\ast}(t) & \rho_{gg}(t) & 0 & 0\\
      0 & 0 & \rho_{ss}(t) & 0\\
      0 & 0 & 0 &\rho_{aa}(t)
    \end{array}\right) ,\label{e9}
\end{eqnarray}
with the density matrix elements evolving as
\begin{eqnarray}
  \rho_{ee}(t) &=& p\,{\rm e}^{-2\gamma t}  ,\nonumber \\
  \rho_{eg}(t) &=& \sqrt{p(1-p)}\,{\rm e}^{-\gamma t}  ,\nonumber \\
  \rho_{ss}(t) &=& p\,{\rm e}^{-\gamma t}\frac{\gamma +\gamma_{12}}{\gamma -\gamma_{12}}
  \left({\rm e}^{-\gamma_{12} t} -{\rm e}^{-\gamma t}\right) ,\nonumber \\
  \rho_{aa}(t) &=&  p\,{\rm e}^{-\gamma t}\frac{\gamma -\gamma_{12}}{\gamma +\gamma_{12}}
  \left({\rm e}^{\gamma_{12} t} -{\rm e}^{-\gamma t}\right)   ,\label{eb}
\end{eqnarray}
subject to conservation of probability
$\rho_{gg}(t)=1-\rho_{ss}(t)-\rho_{aa}(t)-\rho_{ee}(t)$. Note that the
evolution of the density matrix elements is independent of the
dipole-dipole interaction between the atoms, but it is profoundly
affected by the collective damping $\gamma_{12}$. This collective
behavior leads to two distinct timescales of the evolution of the
populations of the symmetric and antisymmetric states, the former much shorter
and the later much longer than that for independent atoms.

Given the density matrix, Eq.~(\ref{e9}), we can now calculate the
concurrence ${\cal C}(t)$ to which we shall later refer as concurrence
in the full sense, and examine the transient dynamics of the
entanglement. First, we find that the square roots of the eigenvalues
of the matrix $R$ are
\begin{eqnarray}
  \sqrt{\lambda_{1,2}(t)} &=& |\rho_{ge}(t)|\pm (\,\rho_{ss}(t)
  +\rho_{aa}(t)\,) , \nonumber \\ 
  \sqrt{\lambda_{3,4}(t)} &=& (\,\rho_{ss}(t)-\rho_{aa}(t)\,) \pm
  \sqrt{\rho_{gg}(t)\rho_{ee}(t)} .\label{eq:roots} 
\end{eqnarray}
from which it is easily verified that for a particular value of the
matrix elements there are two possibilities for the largest
eigenvalue, either $\sqrt{\lambda_{1}(t)}$ or $\sqrt{\lambda_{3}(t)}$.
The concurrence is thus given by
\begin{eqnarray}
  {\cal C}(t) = \max\left\{0,\,{\cal C}_{1}(t),\,{\cal
      C}_{2}(t)\right\} ,\label{eq:concurrence1} 
\end{eqnarray}
with
\begin{eqnarray}
  {\cal C}_{1}(t) &=& 2\,|\rho_{ge}(t)|-(\, \rho_{ss}(t)
  +\rho_{aa}(t)\,)  ,\nonumber\\ 
  {\cal C}_{2}(t) &=&  |\rho_{ss}(t)-\rho_{aa}(t)|
  -2\,\sqrt{\rho_{gg}(t)\rho_{ee}(t)} . \label{eq:altconc} 
\end{eqnarray}
From this it is clear that the concurrence ${\cal C}(t)$ can always be
regarded as being made up of the sum of nonnegative contributions of
the weights ${\cal C}_{1}(t)$ and ${\cal C}_{2}(t)$ associated with
two different classes of entangled states that can be generated in a
two qubit system. From the form of the entanglement weights it is
obvious that ${\cal C}_{1}(t)$ provides a measure of an entanglement
produced by linear superpositions involving the ground $\ket g$ and
the upper $\ket e$ states of the system, whereas ${\cal C}_{2}(t)$
provides a measure of an entanglement produced by a distribution of
the population between the symmetric and antisymmetric states.
Inspection of Eq.~(\ref{eq:altconc}) shows that for ${\cal C}_{1}(t)$
to be positive it is necessary that the two-photon coherence
$\rho_{eg}$ is different from zero, whereas the necessary condition
for ${\cal C}_{2}(t)$ to be positive is that the the symmetric and
antisymmetric states are not equally populated.

We consider first the effect of the collective damping on the sudden
death of an initial entanglement determined by the state (\ref{ea}).
The entanglement weights ${\cal C}_{1}(t)$ and ${\cal C}_{2}(t)$,
which are needed to construct ${\cal C}(t)$ are readily calculated
from Eqs.~(\ref{ea}) and (\ref{eq:altconc}). We see that the system
initially prepared in the state (\ref{ea}) can be entangled according
to the criterion ${\cal C}_{1}$, and the degree to which the system is
initially entangled~is ${\cal C}_{1}(0)= 2\sqrt{p(1-p)}$.

If the atoms radiate independently, $\gamma_{12}=0$, and then we find
from Eq.~(\ref{eb}) that $\rho_{ss}(t) =\rho_{aa}(t)$.  It is clear by
inspection of Eq.~(\ref{eq:altconc}) that in this case ${\cal
  C}_{2}(t)$ is always negative, so we immediately conclude that no
entanglement is possible according to the criterion ${\cal C}_{2}$,
and the atoms can be entangled only according to the criterion~${\cal
  C}_{1}$. The initial entanglement decreases in time because of the
spontaneous emission and disappears at the time
\begin{eqnarray}
  t_{d} =
  \frac{1}{\gamma}\ln\left(\frac{p+\sqrt{p(1-p)}}{2p-1}\right)  ,
  \label{eu} 
\end{eqnarray}
from which we see that the time it takes for the system to disentangle
is a sensitive function of the initial atomic conditions. We note from
Eq.~(\ref{eu}) that the sudden death of the entanglement of
independent atoms is possible only for $p > 1/2$. Since
$\rho_{ee}(0)=p$, we must conclude that the entanglement sudden death
is ruled out for the initially not inverted system.
\begin{figure}[th]
  \includegraphics[height=4cm]{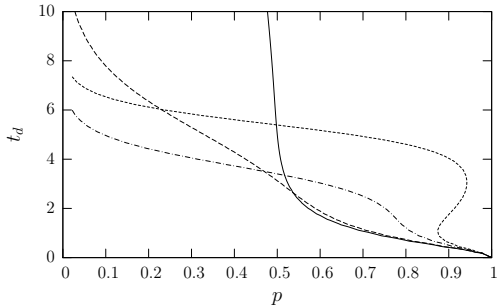}
  \caption{The death time of the entanglement prepared according to
    the criterion ${\cal C}_{1}$ and plotted as a function of $p$ for
    different separations between the atoms: $r_{12}=\lambda$ (solid
    line), $r_{12}=\lambda/3$ (dashed line), $r_{12}=\lambda/6$
    (dashed-dotted line), $r_{12}=\lambda/20$ (dotted line).}
  \label{fig-1}
\end{figure}

For a collective system, when the atoms are close to each other,
$\gamma_{12}\neq 0$, and then the sudden death appears in less
restricted ranges of the parameter~$p$. This is shown in
Fig.~\ref{fig-1}, where we plot the death time as a function of $p$
for several separations between the atoms. We see that the range of
$p$ over which the sudden death occurs increases with decreasing
$r_{12}$, and for small separations the sudden death occurs over the
entire range of $p$. 

The most interesting consequence of the collective damping is the
possibility of the entanglement revival. We now use Eqs.~(\ref{eb})
and (\ref{eq:altconc}) to discuss the ability of the system to revive
entanglement in the simple process of spontaneous emission.
\begin{figure}[th]
  \includegraphics[height=4cm]{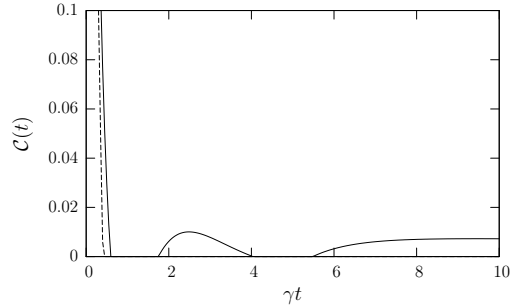}
  \caption{Transient evolution of the concurrence ${\cal C}(t)$ for
    the initial state $\ket{\Psi_{0}}$ with $p=0.9$.  The solid line
    represents ${\cal C}(t)$ for the collective system with the
    interatomic separation $r_{12}=\lambda/20$.  The dashed line shows
    ${\cal C}(t)$ for independent atoms, $\gamma_{12}=0$.}
  \label{fig-2}
\end{figure}
Figure~\ref{fig-2} shows the deviation of the time evolution of the
concurrence for two interacting atoms from that of independent atoms.
In both cases, the initial entanglement falls as the transient
evolution is damped by the spontaneous emission. For independent atoms
we observe the collapse of the entanglement without any revivals.
However, for interacting atoms, the system collapses over a short time
and remains disentangled until a time $t_{r}\approx 1.7/\gamma$ at
which, somewhat counterintuitively, the entanglement revives. This
revival then decays to zero, but after some period of
time a new revival begins. Thus, we see two time intervals (dark
periods) at which the entanglement vanishes and two time intervals at
which the entanglement revives. To estimate the death and revival times,
we use Eqs.~(\ref{eq:altconc}) and (\ref{eb}), and find that for 
$\gamma_{12}\approx \gamma$, the entanglement weight
${\cal C}_{1}(t)$ vanishes at times satisfying the relation
\begin{eqnarray}
  \gamma t\, \exp(-\gamma t) = \sqrt{\frac{1-p}{p}} , \label{ep}
\end{eqnarray}
which for $p>0.88$ has two nondegenerate solutions, $t_{d}$ and
$t_{r}>t_{d}$. The time $t_{d}$ gives the collapse time of the
entanglement beyond which the entanglement disappears. The death zone
of the entanglement continues until the time $t_{r}$ at which the
entanglement revives. Thus, for the parameters of Fig.~\ref{fig-2},
the entanglement collapses at $t_{d}=0.6/\gamma$ and revives at
the time $t_{r}=1.7/\gamma$.

The origin of the dark periods and the revivals of the entanglement
can be understood in terms of the populations of the collective states
and the rates with which the populations and the two-photon coherence
decay. One can note from Eq.~(\ref{eb}) that for short times
$\rho_{aa}(t)\approx 0$, but $\rho_{ss}(t)$ is large. Thus, the
entanglement behavior can be analyzed almost entirely in terms of the
population of the symmetric state and the coherence $\rho_{eg}(t)$.

\begin{figure}[th]
  \includegraphics[height=4cm]{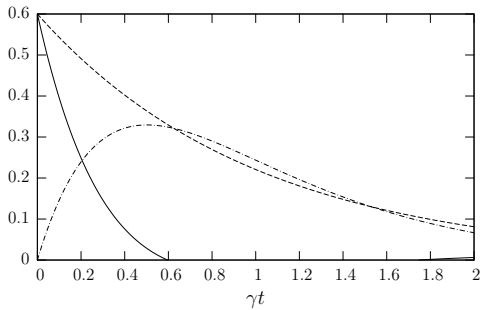}
  \caption{Origin of the first dark period and revival of the
    entanglement of a collective system. The time evolution of the
    coherence $2|\rho_{eg}(t)|$ (dashed line) is compared with the
    evolution of the population $\rho_{ss}(t)$ (dashed-dotted line)
    for the same parameters as in Fig.~\ref{fig-2}. The solid line is
    the time evolution of the concurrence ${\cal C}(t)={\cal C}_{1}(t)$.}
  \label{fig-3}
\end{figure}

Figure~\ref{fig-3} shows the time evolution of ${\cal C}(t)$, the
population $\rho_{ss}(t)$, and the coherence $\rho_{eg}(t)$.  As can
be seen from the graphs, the entanglement vanishes at the time at
which the population of the symmetric state is maximal and remains
zero until the time $t_{r}$ at which $\rho_{ss}(t)$ becomes smaller
than $\rho_{eg}(t)$. We may conclude that the first dark period arises
due to the significant accumulation of the population in the symmetric
state. The impurity of the state of the two-atom system is rapidly
growing and entanglement disappears. 
 
The reason for the occurrence of the first revival, seen in
Fig.~\ref{fig-2} is that the two-photon coherence $\rho_{eg}(t)$
decays more slowly than the population of the symmetric state. Once
$\rho_{ss}(t)$ falls below $2|\rho_{eg}(t)|$, entanglement emerges
again. Thus, the coherence can become dominant again and entanglement
regenerated over some period of time during the decay process. This is
the same coherence that produced the initial entanglement.  Therefore,
we may call the first revival as an "echo" of the initial entanglement
that has been unmasked by destroying the population of the symmetric
state. It is interesting to note that the entanglement revival appears
only for large values of $p$, and is most pronounced for
$p>0.88$. This is not surprising because for $p>1/2$ the system is
initially inverted that increases the probability of spontaneous
emission. 

We have seen that the short time behavior of the entanglement is
determined by the population of the symmetric state of the system.  A
different situation occurs at long times. As it is seen from
Fig.~\ref{fig-2}, the entanglement revives again at longer times and
decays asymptotically to zero as $t\rightarrow \infty$. The second
revival has completely different origin than the first one.  At long
times both $\rho_{ss}(t)$ and $\rho_{eg}(t)$ are almost zero. However,
the population $\rho_{aa}(t)$ is sufficiently large as it accumulates on the time
scale $t=1/(\gamma -\gamma_{12})$ which is very long when
$\gamma_{12}\approx \gamma$. A careful examination of Eq.~(\ref{eb})
shows that ${\cal C}_{1}(t)<0$ at long times, so that the long
time entanglement is determined solely by the weight ${\cal C}_{2}$,
which is negative for short times, and it becomes positive after a
finite time $t_{r_{2}}$ (second revival time) given approximately by
the formula
\begin{eqnarray}
  t_{r_{2}} \approx \frac{1}{\gamma_{12}}\ln\left(\frac{1}{\sqrt{p}}
  \frac{4\gamma}{\gamma -\gamma_{12}}\right)  , \label{eu2}
\end{eqnarray}

It follows from the above analysis and Fig.~\ref{fig-2} that the
entanglement prepared according to the criterion ${\cal C}_{1}$ is
rather short-lived affair compared with a long-lived entanglement
prepared along the criterion ${\cal C}_{2}$. Asymptotically,
the concurrence is equal to the population $\rho_{aa}(t)$.


In summary, we have examined the transient evolution of the
entanglement in a two atom system coupled to the multimode vacuum
field. We have predicted the occurence of dark periods and revivals of
entanglement induced by the irreversible process of spontaneous
emission. The results show that the revivals are independent of the
dipole-dipole interaction between the qubits but crucially depend on
the collective damping. We have shown that this unusual behavior of
the entanglement results from a significant modification of the
spontaneous emission rates of the symmetric and antisymmetric states.


This work was supported in part by the Australian Research Council and
the Polish Ministry of Education and Science grant 1 P03B 064 28.


\begin{thebibliography}{18}
\bibitem{phbk} M. B. Plenio, S. F. Huelga, A. Beige, and P. L. Knight,
  Phys. Rev. A {\bf 59}, 2468 (1999).
\bibitem{ft02} Z. Ficek and R. Tana\'s, Phys. Rep. {\bf 372}, 369
  (2002).
\bibitem{ms04} V. S. Malinovsky and I. R. Sola, Phys. Rev. Lett. {\bf
    93}, 190502 (2004).
\bibitem{sm06} A. Serafini, S. Mancini, and S. Bose, Phys. Rev. Lett.
  {\bf 96}, 010503 (2006).
\bibitem{zg00} S.-B. Zheng and G.-C. Guo, Phys. Rev. Lett. {\bf 85},
  2392 (2000).
\bibitem{os01} S. Osnaghi {\it et al.}, Phys. Rev. Lett. {\bf 87},
  037902 (2001).
\bibitem{dic} R. H. Dicke, Phys. Rev. {\bf 93}, 99 (1954).
\bibitem{le70} R. H. Lehmberg, Phys. Rev. A {\bf 2}, 883 (1970).
\bibitem{ag74} G. S. Agarwal, {\it Quantum Statistical Theories of
    Spontaneous Emission and their Relation to other Approaches},
  edited by G. H\"{o}hler, Springer Tracts in Modern Physics, Vol. 70,
  (Springer-Verlag , Berlin, 1974).
\bibitem{fs05} Z. Ficek and S. Swain, {\it Quantum Interference and
    Coherence: Theory and Experiments} (Springer, New York, 2005).
\bibitem{bd00} G. K. Brennen {\it et al.}, Phys. Rev. A {\bf 61},
  062309 (2000).
\bibitem{ft03} Z. Ficek and R. Tana\'s, J. Mod. Opt. {\bf 50}, 2765
  (2003) and references therein.
\bibitem{ye04} T. Yu and J. H. Eberly, Phys. Rev. Lett. {\bf 93},
  140404 (2004).
\bibitem{ja06} L. Jak\'{o}bczyk and A. Jamr\'{o}z, Phys. Lett {\bf A
    333}, 35 (2004); A. Jamr\'{o}z, quant-ph/0602128 (2006).
\bibitem{woo} W. K. Wootters, Phys. Rev. Lett. {\bf 80}, 2245 (1998).
\end{thebibliography}
\end{document}